\documentclass[runningheads,a4paper]{llncs}

\let\proof\relax
\let\endproof\relax
\usepackage[fleqn]{amsmath}
\usepackage{amssymb,amstext,amsthm}
\usepackage{thmtools}
\usepackage{thm-restate}
\usepackage{hyperref}
\usepackage[dvipsnames]{xcolor}
\usepackage{soul}
\usepackage{mathtools} 
\usepackage{bcprules} 
\usepackage{mdframed} 
\usepackage{lipsum}
\usepackage{multicol}
\usepackage{graphicx}
\usepackage{xspace}
\usepackage{xfrac} 
\usepackage{caption}
\usepackage{enumerate}
\usepackage{lstautogobble}
\usepackage{listings} 
\usepackage{commands} 
\usepackage{natbib}
\usepackage{url}  
\newcommand{\keywords}[1]{\par\addvspace\baselineskip
\noindent\keywordname\enspace\ignorespaces#1}

\lstdefinelanguage{Scala}{
  morekeywords={abstract,case,catch,class,def,%
    do,else,extends,false,final,finally,%
    for,if,implicit,import,match,mixin,%
    new,null,object,override,package,%
    private,protected,requires,return,sealed,%
    super,this,throw,trait,true,try,%
    type,val,var,while,with,yield},
  otherkeywords={=>,<-,<\%,<:,>:,\#,@},
  sensitive=true,
  morecomment=[l]{//},
  morecomment=[n]{/*}{*/},
  morestring=[b]",
  morestring=[b]',
  morestring=[b]"""
}

\begin{document}

\mainmatter 
\title{Mutable WadlerFest DOT}
\titlerunning{Mutable WadlerFest DOT}
\author{Marianna Rapoport \and Ond\v{r}ej Lhot\'ak}
\authorrunning{M. Rapoport, O. Lhot\'ak}
\institute{\{mrapoport, olhotak\}@uwaterloo.ca\\University of Waterloo}
\maketitle

\begin{abstract}
    The Dependent Object Types (DOT) calculus aims to model the essence of Scala,
    with a focus on abstract type members, path-dependent types, and subtyping. Other
    Scala features could be defined by translation to DOT.

    Mutation is a fundamental feature of Scala currently missing in DOT.
    Mutation in DOT is needed not only to model effectful computation and mutation
    in Scala programs, but even to precisely specify how Scala initializes
    immutable variables and fields (\textsf{val}s).

    We present an extension to DOT that adds typed mutable reference cells.
    We have proven the extension sound with a mechanized proof in Coq.
    We present the key features of our extended calculus and its soundness proof,
    and discuss the challenges that we encountered in our search for
    a sound design and the alternative solutions that we considered.
\end{abstract}

\section{Introduction}

Abstract type members, parametric polymorphism,
and mix-in composition are only a few features of Scala's complex type system.
The presence of path-dependent types has made it particularly hard to
understand the interaction between the numerous language components
and to come up with a precise formalization for Scala.
The lack of a theoretical foundation for the language has in turn led to unsound design
choices~\citep{sdts,bugrep,null}.

To model the interaction between Scala's core features soundly,
researchers have worked for over ten years to devise formal
calculi~\citep{nuobj,fs,scalina,fool12,oopsla14,arxiv1,arxiv2,Amin2016,oopsla16,aminphd}.
We refer to the specific calculus of~\cite{Amin2016}
as WadlerFest DOT because several different calculi have used the name DOT.
WadlerFest DOT models the key components of the Scala type system, such as type members, 
path-dependent types, and subtyping.
The eventual intent is to formalize other constituents of the full language, such as classes
and inheritance, by a translation to the core features of DOT.

However, WadlerFest DOT is still lacking some fundamental Scala features, 
one of which is mutation.
Without mutation, it is difficult to model (mutable) variables and fields,
or to reason about side effects in general.

Interestingly, mutation is even necessary to model a sound class
initialization order for \emph{immutable} fields, which are mutated once when they
are initialized.
At the moment, Scala's complex initialization order can lead to programs with unintuitive behaviour of fields~\citep{init}; in particular, current versions of the Scala compiler permit programs in which immutable fields are read before they have been initialized.
In order for the Scala community to discuss alternative designs of the
initialization order, it needs a means to specify candidate designs precisely
and evaluate them formally.
A sound formalization of initialization order, in turn, requires reasoning about overwriting of
class members that first hold a \textsf{null} value from the time that they are
allocated to the time that they are initialized, which is not directly
possible in WadlerFest DOT.

This paper presents the \mdot calculus, which is an extension to WadlerFest DOT with typed mutable references.
To that end, we augment the calculus with a mutable heap and the possibility
to create, update, and dereference mutable memory cells, or locations.
A Scala mutable variable (\textsf{var}) can then be modelled by an 
immutable variable (already included in WadlerFest DOT),
containing a mutable memory cell.
For example, a Scala object
\begin{lstlisting}[language=Scala]
object O {
  val x = 1
  var y = 2
}
\end{lstlisting}
can be represented in mutable-DOT pseudocode as follows:%
\footnote{The Scala type system is nominal while WadlerFest DOT is (mostly) structural.
Therefore, the Scala example assigns the object a name, while WadlerFest DOT does not.}
\begin{lstlisting}
new {this: {x: Int} /*$\wedge$*/ {y: Ref Int}} // structural type of object
  {x = 1} /*$\wedge$*/ {y = ref 2 Int}         // definitions in object body
\end{lstlisting}
An unusual characteristic of our heap implementation is that it maps
locations to variables instead of values.
This design choice is induced by WadlerFest DOT's type system, which disallows
subtyping between recursive types.
We show how, as a result, storing \textit{values} on the heap would significantly
limit the expressiveness of our calculus, and explain the correctness
of storing \textit{variables} on the heap.

It is not surprising that adding mutable references to a DOT calculus
is \emph{possible}.
Indeed, in an update to their technical report,
\citet{arxiv2} report that they added them to
a different DOT calculus with a big-step semantics.

WadlerFest DOT is well suited as a basis for future extension,
both to specify existing higher-level Scala features by translation
to a core calculus, and to formally explore new proposed extensions
to Scala.
It comes with a soundness proof
formalized and verified in Coq.
WadlerFest DOT is simpler than the other full DOT
calculi, and its
semantics is
small-step, so the soundness proof is based on the
familiar approach of progress and preservation~\citep{felleisen}.

The contributions of this paper are:
\begin{itemize}
  \item We define an operational semantics and type system for \textit{\mdot}, an extension of the small-step WadlerFest DOT calculus with mutable references.
  \item We provide a mechanized type safety proof in Coq, in the form of an extension of the original WadlerFest DOT proof, which is suitable to be used for extensions
      of WadlerFest DOT that require mutation.%
\footnote{The mechanized proof can 
be found in our fork of the WadlerFest DOT proof repository (see 
\url{https://github.com/amaurremi/dot-calculus}).}
  \item We discuss the challenges that we encountered in adding
      mutation to WadlerFest DOT, and the design choices that
      we made to overcome them. We conjecture that this reported experience
      will be helpful for adding mutation to other DOT
      calculi, including the OOPSLA DOT, which is to appear~\citep{oopsla16}.
\end{itemize}

The rest of the paper is organized as follows.
Section~\ref{sec:dot} is a brief introduction to the original WadlerFest DOT calculus.
Section~\ref{sec:mutation} presents the mutable DOT calculus \mdot,
and Section~\ref{sec:safety} outlines its type-safety proof.
We discuss \mdot's design in Section~\ref{sec:discussion} and
present an example of using mutable references in \mdot
in Section~\ref{sec:example}.
Related work is discussed in~Section~\ref{sec:related}.

\section{The WadlerFest DOT Calculus}\label{sec:dot}
We introduce the features of the WadlerFest DOT calculus through an example.
Suppose we want to keep track of fish that live in aquariums.
In Scala, we could write:
  \listingNoCapt{Scala}{aquarium.scala}
This program lets us add a fish \textsf{gf} to the \textsf{goldfish} 
aquarium:

\begin{lstlisting}[language=Scala]
val gf: Goldfish = ...
addFish(goldFish, gf)
\end{lstlisting}
but it will result in a type error when trying to add \textsf{gf} to the \textsf{piranha}
aquarium:

\begin{lstlisting}[language=Scala]
addFish(piranhas, gf)
// type error: expected: piranhas.Fish, actual: goldFish.Fish
\end{lstlisting}
The reason the goldfish is protected from the piranhas is that the 
type \textsf{Fish} is \textit{path dependent}, i.e. specific to the
run-time \textsf{Aquarium} object that the fish belongs to.
This allows the \textsf{addFish} method to guarantee at compile time that an aquarium
\textsf{a} accepts only fish of type \textsf{a.Fish}.

The syntax, reduction semantics, and type system of the WadlerFest DOT calculus can be seen
in Figures~\ref{fig:synt}, \ref{fig:red}, \ref{fig:typing}, and~\ref{fig:subtyping}.
The shaded parts are our mutation-related changes and can be ignored for now.

As a first attempt to define \textsf{Aquarium} in WadlerFest DOT, we can make it an
\textit{intersection} of two types:

\begin{lstlisting}
{ Aquarium = {Fish: /*$\bot..\top$*/} /*$\wedge$*/ {fish: List} }
\end{lstlisting}
The first type, $\tTypeDec {\mathsf{Fish}} \bot \top$, declares a \textit{type 
member} \textsf{Fish} with lower bound $\bot$ (\textsf{Nothing}) and upper bound
$\top$ (\textsf{Any}).
The second type, $\tFldDec {\mathsf{fish}} {\mathsf{List}}$, is a \textit{field
declaration} of type \textsf{List} that represents the list of \textsf{fish} in the aquarium.
The type \textsf{List} is assumed to be defined in a library and contains a type member
\textsf A for list elements.

A problem with the current \textsf{Aquarium} implementation is that it does not say
that the type of elements in the \textsf{fish} list should be \textsf{Fish}.
More specifically, the list elements should have the \textsf{Fish} type of the
\textsf{Aquarium} \textit{runtime object} to which the list belongs.
To let the \textsf{Aquarium} type refer to its own runtime object \textsf a, we make 
\textsf{Aquarium} a \textit{recursive type}:

\begin{lstlisting}
{ Aquarium = /*$\mu$*/(a: {Fish: /*$\bot..\top$*/} /*$\wedge$*/
                   {fish: List /*$\wedge$*/ {A: a.Fish..a.Fish}}}
\end{lstlisting}

Here, to express that the type \textsf{Fish} should belong to the object \textsf a, we 
use the \textit{type selection} \textsf{a.Fish}.
The type \textsf{a.Fish} is then used as a refinement of \textsf{List}'s element type
\textsf A.
In this way, the list can contain only the fish that are allowed in 
the aquarium \textsf a.

We can now define \textsf{addFish} as a function that takes an aquarium \textsf a and a 
fish \textsf f of type \textsf{a.Fish}, and creates a new aquarium \textsf{a2}:

\begin{lstlisting}
{ addFish = /*$\lambda$*/(a: aq.Aquarium)./*$\lambda$*/(f: a.Fish).
    /*$\nu$*/(a2: aq.Aquarium /*$\wedge$*/{ Fish: a.Fish..a.Fish }) {
      Fish = a.Fish
      fish = ... }}
\end{lstlisting}
The construct $\tNew x T d$ creates a new object of type $T$ with a
self-variable $x$ and definitions $d$.
In this case, the definitions are used to initialize the \textsf{Fish} type and
\textsf{fish} list of the new aquarium. 
The \textsf{Fish} type is assigned \textsf{a.Fish}.
The new \textsf{fish} list needs to append the fish \textsf f to the old \textsf{a.fish}
list.

To be able to add an element to a list, we need access to an append method, which we
will get from \textsf{List}.
Suppose that the \textsf{List} type is defined in a \textsf{collections} library.
It can be defined as a recursive type $\mu(\mathsf{list}\colon\dots)$ that declares
an element type \textsf A and an \textsf{append} function.
\textsf{append} takes a parameter \textsf a of the element type \textsf{list.A} and
returns a \textsf{List} of elements that are subtypes of \textsf{a.A}:

\begin{lstlisting}
let collections = /*$\nu$*/(col: 
  {List: /*$\mu$*/(list: ({A: /*$\bot..\top$*/} /*$\wedge$*/ 
        {append: /*$\forall$*/(a: list.A)(col.List /*$\wedge$*/ {A: /*$\bot$*/..a.A}))}) ...
in ...
\end{lstlisting}
With an \textsf{append} method on \textsf{List}s, we can fully implement the
\textsf{addFish} method. The field \textsf{a2.fish} should be defined as 
\textsf{a.fish.append(f)}. 
However, since WadlerFest DOT uses administrative normal form (ANF), before performing any operations on terms, we have to
bind the terms to variables:
\begin{lstlisting}
fish = let oldFish = a.fish         in
       let append  = oldFish.append in
       append f
\end{lstlisting}

For better readability, we introduce the following abbreviations 
(similar ones are used in the WadlerFest DOT paper):
{\setlength{\mathindent}{0cm}\begin{align*}
  \set{A} \equiv&  \set{A\colon\bot..\top}             &&& t\,u    \equiv&\,  \mlet x=t \ \tin\\
  \set{A\colon T}  \equiv&  \set{A\colon T..T}         &&&               &\,  \tLet y u {x\,y}\\
  \set{A<: T}      \equiv&  \set{A\colon\bot..T}       &&& t.L     \equiv&\,  \tLet x t {x.L}\\
  \set{D_1; D_2}   \equiv&  \set{D_1}\wedge\set {D_2}  &&& \nu(x)d\colon T  \equiv&\,  \nu(x\colon T)d
\end{align*}}
where $D_1\co D_2$ are declarations or definitions of either fields or types,
and $L$ is a label of a type or field.

With those abbreviations, the full aquarium program example looks as follows:

  \listingNoCapt{ML}{aquarium.dot}
  
\section{Mutation in WadlerFest DOT}\label{sec:mutation}

In this section, we present \mdot, our extension of the WadlerFest DOT calculus with mutable references. The changes in the syntax are inspired by Pierce's extension of
the simply-typed lambda calculus with
references~\citep[Chapter~13]{Pierce:2002:TPL:509043}.

Throughout this paper, we highlight the changes necessary to convert WadlerFest DOT into \mdot in 
grey.

\subsection{Abstract syntax}
To support mutation, we augment the WadlerFest DOT syntax with \textit{references} that point to \textit{mutable memory cells}, or \textit{locations}, as shown in Figure~\ref{fig:synt}. 
\begin{wide-rules}
  \begin{multicols}{2}

$\andalso\andalso\quad$\tnew{\parbox{0.36\textwidth}{
    \begin{flalign}
      \sto&\Coloneqq     \tag*{\textbf{Store}}\\
        &\varnothing     \tag*{empty store}\\
        &\extendSto l x  \tag*{extended or updated store}
    \end{flalign}
}}
    \begin{flalign}
  x,\,y,\,z            \tag*{\textbf{Variable}}\\
  a,\,b,\,c            \tag*{\textbf{Term member}}\\
  A,\,B,\,C            \tag*{\textbf{Type member}}\\
  S,\,T,\,U&\Coloneqq  \tag*{Type}\\
    &\top              \tag*{top type}\\
    &\bot              \tag*{bottom type}\\
    &\tFldDec a T      \tag*{field declaration}\\
    &\tTypeDec A S T   \tag*{type declaration}\\
    &x.A               \tag*{type projection}\\
    &\tAnd S T         \tag*{intersection}\\
    &\tRec x T         \tag*{recursive type}\\
    &\tForall x S T    \tag*{dependent function}\\
    &\new{\Ref T}      \tag*{\new{\text{reference type}}}\\
  v&\Coloneqq          \tag*{\textbf{Value}}\\
   &\tNew x T d        \tag*{object}\\
   &\tLambda x T t     \tag*{lambda}\\
   &\new{l}            \tag*{\new{\text{location}}}\\
  s,\,t,\,u&\Coloneqq  \tag*{\textbf{Term}}\\
   &x                  \tag*{variable}\\
   &v                  \tag*{value}\\
   &x.a                \tag*{selection}\\
   &x\,y               \tag*{application}\\
   &\tLet x t u        \tag*{let binding}\\
   &\new{\tRef x T}    \tag*{\new{\text{reference}}}\\
   &\new{!x}           \tag*{\new{\text{dereferencing}}}\\
   &\new{x\coloneqq y} \tag*{\new{\text{assignment}}}\\
  d&\Coloneqq          \tag*{\textbf{Definition}}\\
   &\set{a=t}          \tag*{field definition}\\
   &\set{A=T}          \tag*{type definition}\\
   &\tAnd d {d'}       \tag*{aggregate definition}
    \end{flalign}

  \end{multicols}
  \caption{Abstract syntax of \mdot}  
  \label{fig:synt}
\end{wide-rules}
Locations are a new kind of value that is added to the syntax, and are denoted as~$l$.
The syntax comes with three new terms to support the following reference operations:
\begin{itemize}
  \item $\tRef x T$ \textit{creates} a new reference of type $T$ in the store
      and initializes it with the variable $x$.
      Section~\ref{sec:refs} explains why reference expressions need to contain a 
      declared type $T$, unlike the references in Pierce's book.
  \item $!x$ \textit{reads} the contents of a reference~$x$.
  \item $\tAsgn x y$ \textit{updates} the contents of a reference~$x$ with the variable 
        $y$.
\end{itemize}
The operations that create, read, and update references operate on variables,
not arbitrary terms, in order to preserve ANF.

Newly-created references become \textit{locations}, or memory addresses, denoted as~$l$.
Locations are stored in the \textit{store}, denoted as $\sto$, which serves as a heap.

The store is a \textit{map} from locations to \textit{variables}.
This differs from the common definition of a store, which maps locations to values.
We discuss the motivation for this design choice in Section~\ref{sec:heapmotivation}.
In order to preserve the commonly expected intuitive behaviour of a store,
we must be sure that while a variable is in
the store, it does not go out of scope or change its value.
We show this in Section~\ref{sec:heapcorrectness}.

Updating a store $\sto$ that contains a mapping $l\mapsto x$ with a new mapping
$l\mapsto y$ overwrites $x$ with $y$:
\[
  (\sto[l\to x])(l')=
    \begin{cases}
      x        &\text{if $l=l'$}\\
      \sto(l') &\text{otherwise}.
    \end{cases}
\]

Locations are typed with the reference type $\Ref T$. 
The underlying type $T$ indicates that the location stores variables of type $T$.

To write concise \mdot programs, we extend the abbreviations from 
Section~\ref{sec:dot} with the following rules:
\begin{align*}
  \tRef t T   &\equiv  \tLet x t {\tRef x T}\\
  \tAsgn t u  &\equiv  \tLet x t {\tLet y u {\tAsgn x y}}\\
  !t         &\equiv  \tLet x t {!x}\\
  t;\ u       &\equiv  \tLet x t u
\end{align*}
\subsection{Reduction rules}

Since the meaning of an \mdot term depends on the store contents, we represent a program 
state as a tuple $\stDft t$, denoting a term~$t$ that can point to memory contents in the 
store~$\sto$.

The new reduction semantics is shown in~Figure~\ref{fig:red}:
\begin{itemize}
  \item A newly created reference $\tRef x T$ reduces to a fresh location with an updated 
        store that maps $l$ to $x$ (\rn{Ref}).
  \item Dereferencing a variable using $!x$ is possible if $x$ is bound to a location $l$ 
        by a \textsf{let} expression.
        If so, $!x$ reduces to $\sto(l)$, the variable stored at location $l$ (\rn{Deref}).
  \item Similarly, if $x$ is bound to $l$ by a \textsf{let}, then the assignment 
        operation $\tAsgn x y$ updates the store at location $l$ with the variable $y$ (\rn{Store}).
\end{itemize}

Programs written in the \mdot calculus generally do not contain explicit
location values in the original program text. Locations are included as values in
the \mdot syntax only because terms such as $\tRef x T$ will evaluate to fresh
locations during reduction.

\begin{wide-rules}

\begin{flalign}
  e&\Coloneqq []\ |\ \tLet x {[]} t\ |\ \tLet x v e
                       \tag*{\textbf{Evaluation context}}
\end{flalign}

\infrule[Term]
  {\reductionN \sto t {\sto'} {t'}}
  {\reductionN {\sto} {e[t]} {\sto'}{e[t']}}

\infrule[Apply]
  {v=\tLambda z T t}
  {\reductionDftN {\tLet x v {e[x\, y]}} {\tLet x v {e[\tSubst z y t]}}}

\infrule[Project]
  {v=\tNew x T {\dots {\set{a=t} \dots}}}
  {\reductionDftN {\tLet x v {e[x.a]}} {\tLet x v {e[t]}}}

\infax[Let-Var]
  {\reductionDftN {\tLet x y t} {\tSubst x y t}}

\infax[Let-Let]
  {\reductionDftN {\tLet x {\tLet y s t} u} {\tLet y s {\tLet x t u}}}

\newruletrue

\infrule[Ref]
  {l\notin\dom\sto}
  {\reductionN \sto {\tRef x T} {\extendSto l x} l}

\newruletrue

\infax[Store]
  {\reductionN \sto {\tLet x l {e[\tAsgn x y]}} {\extendSto l y} {\tLet x l {e[y]}}}

\newruletrue

\infrule[Deref]
  {\sto(l)=y}
  {\reductionDftN {\tLet x l {e[!x]}} {\tLet x l {e[y]}}}

    \caption{Reduction rules for \mdot}
    \label{fig:red}
  \end{wide-rules}

The remaining rules are the WadlerFest DOT evaluation rules, with
the only change that they pass along a store.

\subsection{Type rules}\label{sec:typing}

The \mdot typing rules, depicted in~Figure~\ref{fig:typing}, depend on
a \textit{store typing}~$\S$ in addition to a type environment~$\G$.
A store typing maps locations to the types of the variables that they store.

\begin{wide-rules}
  \begin{multicols}{2}

\begin{flalign}
  \G&\Coloneqq \varnothing\ |\ \extendG x T
                       \tag*{\textbf{Type environment}}\\
  \new{\strut \S}&\new{\strut \Coloneqq \varnothing\ |\ \extendS l T}
                       \tag*{\textbf{\tnew{\strut Store typing}}}
\end{flalign}

\infrule[Var]
  {\G(x)=T}
  {\typDftN x T}
  
\newruletrue
\infrule[Loc]
  {\S(l)=T}
  {\typDft l {\Ref T}}
\newrulefalse

\infrule[All-I]
  {\typN {\extendG x T} \S t U
    \andalso
    x\notin\fv T}
  {\typDftN{\tLambda x T t}{\tForall x T U}}

\infrule[All-E]
  {\typDftN x {\tForall z S T}
    \andalso
    \typDftN y S}
  {\typDftN {x\, y} {\tSubst z y T}}

\infrule[\{\}-I]
  {\typN{\extendG x T} \S d T}
  {\typDftN {\tNew x T d} {\tRec x T}}

\infrule[\{\}-E]
  {\typDftN x {\tFldDec a T}}
  {\typDftN {x.a} T}
  
\infrule[Let]
  {\typDftN t T
      \\
    \typN {\extendG x T} \S u U
    \andalso
    x\notin\fv U}
  {\typDftN {\tLet x t u} U}

\infrule[Rec-I]
  {\typDftN x T}
  {\typDftN x {\tRec x T}}

\infrule[Rec-E]
  {\typDftN x {\tRec x T}}
  {\typDftN x T}

\infrule[\&-I]
  {\typDftN x T
    \andalso
    \typDftN x U}
  {\typDftN x {\tAnd T U}}

\infrule[Sub]
  {\typDftN t T
    \andalso
    \subDftN T U}
  {\typDftN t U}

\infrule[Fld-I]
  {\typDftN t T}
  {\typDftN {\set{a=t}} {\tFldDec a T}}

\infax[Typ-I]
  {\typDftN {\set{A=T}} {\tTypeDec A T T}}

\infrule[AndDef-I]
  {\typDftN {d_1} {T_1}
    \andalso
    \typDftN {d_1} {T_2}
    \\
    \dom{d_1},\,\dom{d_2}\text{ disjoint}}
  {\typDftN {\tAnd {d_1} {d_2}} {\tAnd {T_1} {T_2}}}

\newruletrue

\infrule[Ref-I]
  {\typDftN x T}
  {\typDftN {\tRef x T}{\Ref T}}

\infrule[Ref-E]
  {\typDftN x {\Ref T}}
  {\typDftN {\ !x} T}

\infrule[Asgn]
  {\typDftN x {\Ref T}
    \andalso
    \typDftN y T}
  {\typDftN {\tAsgn x y} T}

  \end{multicols}
  
  \caption{Type rules for \mdot}
  \label{fig:typing}

\end{wide-rules}

The store typing spares us the need to re-typecheck locations
and allows to typecheck cyclic references~\citep{Pierce:2002:TPL:509043}.

As an example, the following \mdot program cannot be easily typechecked
without an explicit store typing (using only the runtime store and the type environment):
\newcommand{\exFOne}{\code{id}}
\newcommand{\exA}{r} 
\newcommand{\exFTwo}{\exFOne'}
\newcommand{\exAPrime}{\exFOne_{\text{der}}}
\begin{align*}
  p=\begin{pmatrix*}[l]
    \begin{matrix*}[l]
      \code{let }  &\exFOne  &=\tLambda x {\top} x
        &\code{in}\\
      \code{let }  &\exA     &=\tRef \exFOne {(\top\to\top)}
        &\code{in}\\
      \code{let }  &\exFTwo  &=\tLambda x {\top} {(!\exA)\,x}
        &\code{in}\\
    \end{matrix*}\\
    \exA \coloneqq \exFTwo
  \end{pmatrix*}
\end{align*}

Starting with an empty store, after two reduction steps we get 
\[
  \st\varnothing p\redt\st{\set{l\to \exFTwo}} {p'},
\]
where
\begin{align*}
  p'=\begin{pmatrix*}[l]
    \begin{matrix*}[l]
      \code{let }  &\exFOne  &=\tLambda x {\top} x
        &\code{in}\\
      \code{let }  &\exA     &=\new l
        &\code{in}\\
      \code{let }  &\exFTwo  &=\tLambda x {\top} {(!\exA)\,x}
        &\code{in}\\
    \end{matrix*}\\
    \new\exFTwo
  \end{pmatrix*}
\end{align*}
We would see by looking into the store that to typecheck the location $l$, we needed to 
typecheck $\exFTwo$.
$\exFTwo$ depends on $\exA$, which in turn refers to the location $l$, creating a cyclic 
dependency.

We therefore augment our typing rules with a store typing, allowing us 
to typecheck each location once and for all, at the time of a reference creation.
In the example, we would know
that $l$ is mapped to $(\top\to\top)$ from the \code{let}-binding of $\exA$
and remember this typing in $\S$.
To express that a term $t$ has type $T$ under the type environment $\G$ and store typing 
$\S$, we write $\typDft t T$.

The typing rules for \mdot are shown in~Figure~\ref{fig:typing}.
The WadlerFest DOT rules are intact except that all typing derivations carry a store 
typing.
The new rules related to mutable references are as follows:
\begin{itemize}
  \item We typecheck locations by looking them up in the store typing.
        If, according to $\S$, a location $l$ stores a variable of type $T$,
        then $l$ has type $\Ref T$ (\rn{Loc}).
  \item A newly created reference $\tRef x T$ can be initialized with the
        variable $x$ if $x$ has type $T$.
        In particular, if $x$'s precise type $U$ is a subtype of $T$, 
        then $x$ has type $T$ by \rn{Sub}, so we can
        still create a $\tRef x T$ (\rn{Ref-I}).
  \item Conversely, dereferencing a variable of a reference type $\Ref T$ yields 
        the type~$T$~(\rn{Ref-E}).
  \item Finally, if $x$ is a reference of type $\Ref T$, we are allowed
        to store a variable $y$ into it if $y$ has type $T$.
        To avoid the need to add a \code{Unit} type to the type system,
        we define an assignment $\tAsgn x y$ to reduce to $y$, so the type 
        of the assignment is $T$ (\rn{Asgn}).
\end{itemize}

\begin{wide-rules}
  \begin{multicols}{2}
    
\infax[Top]
  {\subDftN T \top}
\infax{}
\infax[Bot]
  {\subDftN \bot T}
\infax{}
\infax[Refl]
  {\subDftN T T}
\infax{}
\infrule[Trans]
  {\subDftN S T
    \andalso
    \subDftN T U}
  {\subDftN S U}
\infax{}
\infax[And$_1$-$<:$]
  {\subDftN {\tAnd T U} T}
\infax{}
\infax[And$_2$-$<:$]
  {\subDftN {\tAnd T U} U}
\infax{}
\infrule[$<:$-And]
  {\subDftN S T
    \andalso
    \subDftN S U}
  {\subDftN S {\tAnd T U}}
  \infax{}
\infrule[$<:$-Sel]
  {\typDftN x {\tTypeDec A S T}}
  {\subDftN S {x.A}}
\infax{}
\infrule[Sel-$<:$]
  {\typDftN x {\tTypeDec A S T}}
  {\subDftN {x.A} T}
\infax{}
\infrule[$<:$-And]
  {\subDftN S T
    \andalso
    \subDftN S U}
  {\subDftN S {\tAnd T U}}
\infax{}
\infrule[Fld-$<:$-Fld]
  {\subDftN T U}
  {\subDftN {\tFldDec a T} {\tFldDec a U}}
\infax{}
\infrule[Typ-$<:$-Typ]
  {\subDftN {S_2} {S_1}
    \\
    \subDftN {T_1} {T_2}}
  {\subDftN {\tTypeDec A {S_1} {T_1}} {\tTypeDec A {S_2} {T_2}}}
\infax{}
\infrule[All-$<:$-All]
  {\subDftN {S_2} {S_1}
    \\
    \subN {\extendG x {S_2}} \S {T_1} {T_2}}
  {\subDftN {\tForall x {S_1} {T_1}} {\tForall x {S_2} {T_2}}}
\infax{}
\newruletrue
\infrule[\tnew{Ref-Sub}]
  {\subDft T U \andalso \subDft U T}
  {\subDft {\Ref T} {\Ref U}}
\newrulefalse
  \end{multicols}
  
  \caption{Subtyping rules for \mdot}
  \label{fig:subtyping}
\end{wide-rules}

\subsection{Subtyping rules}\label{sec:subtyping}

The subtyping rules of \mdot include an added store typing, and a subtyping rule for references.
The rules are shown in Figure~\ref{fig:subtyping}.

Subtyping between reference types is invariant: usually, $\Ref T<:\Ref U$ if and 
only if $T=U$.
Invariance is required because reference types need to be (i)~covariant for reading, or dereferencing,
and (ii)~contravariant for writing, or assignment.

However, in WadlerFest DOT, co- and contra-variance between types does not imply type equality:
the calculus contains examples of types that are
not equal, yet are equivalent with respect to subtyping. For example,
for any types $T$ and $U$, $\tAnd T U <: \tAnd U T <: \tAnd T U$. Yet, $\tAnd T U\ne\tAnd U T$.
Therefore, subtyping between reference types requires both covariance and contravariance:
\newruletrue
\infrule[Ref-Sub]
  {\subDft T U \andalso \subDft U T}
  {\subDft {\Ref T} {\Ref U}}
\newrulefalse

\section{Type Safety}\label{sec:safety}
In this section, we outline the soundness proof of \mdot as an extension
of the WadlerFest DOT soundness proof~\citep{Amin2016}.

To formulate the progress theorem, Amin~et~al. introduce the notion of an \textit{answer}:
\[
  n\Coloneqq x\mid v\mid\tLet x v n.
\]

The type safety of \mdot is expressed as an extension to the WadlerFest DOT progress and
preservation theorems. The theorems include a store~$\sto$ and a store typing~$\S$,
where~$\S$, unlike~$\G$, can be non-empty.

\begin{theorem}[Progress]\label{thm:progress}
  If $\typN\varnothing\S t T$, then either $t$ is an answer, or \tnew{for} \tnew{any store 
  $\sto$ such that $\wt{\varnothing}\S\sto$}, there is a term $t'$ and \tnew{a store 
  $\sto'$} such that $\reductionN\sto t {\sto'} {t'}$.
\end{theorem}

\begin{theorem}[Preservation]\label{thm:preservation}
  If 
  \begin{itemize}
    \item $\typN\varnothing\S t T$
    \item $\new{\wt{\varnothing}\S\sto}$
    \item $\reductionN\sto t {\sto'} {t'}$,
  \end{itemize}
  then \tnew{for some $\S'\supseteq\S$},
  \begin{itemize}
    \item $\typN\varnothing{\S'}{t'}T$
    \item $\new{\wt{\varnothing}{\S'}{\sto'}}$.
  \end{itemize}
\end{theorem}

Below we describe how to extend the WadlerFest DOT proof to prove \mdot soundness.
Our paper comes with a mechanized Coq proof, which is also an extension of the WadlerFest DOT proof.
The Coq proof can be found in our fork of the WadlerFest DOT proof repository:
\begin{center}\url{https://github.com/amaurremi/dot-calculus}\end{center}

\subsection{Main ideas of the WadlerFest DOT soundness proof}
We start by introducing the key ideas of the WadlerFest DOT proof.
We will later show how to adapt them to prove \mdot type safety.

\paragraph{Bad bounds}
One of the challenges of proving DOT sound is the problem of ``bad 
bounds''~\citep{fool12}. For every pair of arbitrary types $T$ and
$U$, there exists an environment $\G$ such that $\G \vdash T<:U$.
Specifically, when type checking the function
$\tLambda y {\tTypeDec A T U} t$, the body $t$ of the function
is type checked in a type environment $\G$ in which
$\G(y) = \tTypeDec A T U$.
Then
$\G \vdash T<:y.A$ and
$\G \vdash y.A<:U$, so
$\G \vdash T<:U$
(using (\rn{$<:$-Sel}), (\rn{Sel-$<:$}), and (\rn{Trans})).
In particular, if $T$ and $U$ are chosen as $\top$ and $\bot$, respectively,
then we get $\G \vdash \top<:\bot$. Since every type is a subtype of $\top$
and a supertype of $\bot$, this means that \emph{all} types become equivalent
with respect to subtyping in this environment.
Thus, if arbitrary type environments were possible,
the type system would collapse, all types would be subtypes of each other,
and types would give us no information about terms.

To avoid bad bounds, Amin~et~al.\ observe that such a type environment
cannot occur for an evaluation context during a concrete execution of the program.
Specifically, if $t'$ is a subterm of some term $t$,
then the type checking rules for $\emptyset \vdash t : T$ require the subterm $t'$
to be type checked in some specific environment $\G$ (i.e. $\G \vdash t' : T'$).
If there is some variable $y$ such that 
$\G \vdash y : \tTypeDec A T U$,
then $y$ must be bound somewhere in $t$ outside of $t'$.
If $t'$ is in an evaluation context of $t$ 
(i.e. $t = e[t']$),
then
the syntactic definition of an evaluation context ensures that $y$ can only
be bound to a \emph{value} by a binding of the form
$\tLet y v u$.
Since $v$ is a value, it binds $A$ with some specific type $S$, so its
type is $\tTypeDec A S S$ by (\rn{Typ-I}).

\paragraph{Precise typing}
In order to reason about ``good'' bounds, the paper introduces the \textit{precise typing} 
relation, denoted as $\vdash_!$.
A precise typing derivation is allowed to use only a subset of WadlerFest DOT's type rules, so as to 
eliminate the rules that can lead to non-equal lower and upper type bounds.

The typing derivation of a value is said to be precise if its root is either
(\rn{$\{\}$-I}) (typing an object) or (\rn{All-E}) (typing an abstraction).%
\footnote{We omit the definition of precise typing for variables because our proof 
modifications hardly affect it. Please refer to Amin~et~al.'s paper for the full 
definition.}
Since the only other rule that could complete a value's typing derivation is subsumption 
(\rn{Sub}), precise typing computes a value's most specific type.

\begin{wide-rules}
  
\begin{flalign}
  \sta&\Coloneqq \varnothing\ |\ \extendSta x v
                       \tag*{\textbf{Stack}}\\
  \new{\strut \sto}&\new{\strut \Coloneqq \varnothing\ |\ \extendSto l x}
                       \tag*{\new{\strut \textbf{Store}}}
\end{flalign}

\infrule[Project]
  {\sta(x)=\tNew x T {\dots\{a = t\}\dots}}
  {\reductionDft
    {x.a}
    t}

\infrule[Apply]
  {\sta(x)=\tLambda z T t}
  {\reductionDft
    {x\,y}
    {\tSubst z y t}}
  
\infax[Let-Var]
  {\reductionDft
    {\tLet x y t}
    {\tSubst x y t}}

\infax[Let-Value]
  {\reductionFull
    {\sstDftN{\tLet x v t}}
    {\sstN{\extendSta x v} \sto t}}

\infrule[Ctx]
  {\reductionFull
    {\sstDftN t}
    {\sstN {\sta'} {\sto'} {t'}}}
  {\reductionFull
    {\sstDftN{\tLet x t u}}
    {\sstN{\sta'}{\sto'}{\tLet x {t'} u}}}

\newruletrue

\infrule[Ref]
  {l\notin\dom\sto}
  {\reductionFull
    {\sstDft{\tRef x T}}
    {\sst \sta {\extendSto l x} l}}

\infrule[Store]
  {\sta(x)=l}
  {\reductionFull
    {\sstDft{\tAsgn x y}}
    {\sst \sta {\extendSto l x} y}}

\infrule[Deref]
  {\sta(x)=l
    \andalso
    \sto(l)=y}
  {\reductionDft
    {!x}
    y}
    
    \caption{Reduction rules for \mdot in which the evaluation context is replaced with a stack for $\code{let}$ bindings. The underlying DOT reduction rules are taken from the Coq proof that accompanies the paper of~\citet{Amin2016}.}
    \label{fig:red2}
\end{wide-rules}

\paragraph{Stack-based reduction rules}
To make more explicit the evaluation order of subterms in evaluation contexts,
Amin~et~al.\ define an equivalent reduction semantics without evaluation contexts that uses a
variable environment as syntactic sugar for a series of
let bindings whose expressions have already been evaluated to values.
In the WadlerFest DOT paper, the variable environment is called a \textit{store}.
We call it a \textit{stack}, and reserve the term \textit{store}
for the mutable heap.
The stack-based reduction relation (including our \mdot extensions)
is shown in Figure~\ref{fig:red2}.
As soon as a \textsf{let}-bound variable $x$ evaluates to a value $v$, the binding
$x\mapsto v$ is moved onto the stack $\sta$ using the Rule (\rn{Let-Value}). 

Although the stack and store appear similar, they have important differences.
A stack needs to support only the lookup and append operations, since we never
perform updates on the stack.
A stack also needs to have a notion of order since values can
refer to variables defined earlier in the stack.
A store on the other hand needs to support appending \textit{and} overwriting 
locations with different variables. 
The store does not need to be ordered because variables cannot refer to locations.
For those reasons, in the Coq formalization of the soundness proof,
the stack is represented as a list, and the store as a map 
data structure.

The stack is an optional element of the calculus, while the store is necessary.
A stack is just syntactic sugar for \textsf{let}-bindings:
$t$ and $\sta\mid t'$ can be alternative, but equivalent ways of writing the same
term.
However, there is no way to write a term $\stDft t$ as just a $t$.
Consequently, we can write $\stDft t$ and $\sst\sta\sto{t'}$ as equivalent 
programs.

\paragraph{Stack correspondence}

The precise type of a value $v$ cannot have bad bounds because to every
type member $A$ that $v$ defines, it assigns a concrete type $T$,
so the upper and lower bounds in the precise type of $v$ must both be $T$:
$\G\vdash_! v\colon \{A : T .. T\}$.
A type environment $\G$ is said to \textit{correspond} to a stack $\sta$
(written $\G\wfsym\sta$) if it assigns to every variable $x$ the
precise type of the corresponding value $\sta(x)$.
In such a type environment, variables cannot have type members with bad bounds.

\paragraph{Possible types}
To prove the Canonical Forms Lemmas, the WadlerFest DOT paper introduces the set of \textit{possible 
types} $\pt \G x v$. Informally, this set is defined to contain the types that
one would expect $x$ to have if it is bound to $v$, in the absence of bad
bounds in $\G$.
The paper then proves that if $\G\wfsym\sta$, then all of the types
$T$ such that $\G\vdash x\colon T$ are actually included in
$\pt \G x \sta(x)$.

\subsection{Adjusting Definitions to \mdot}
To extend the WadlerFest DOT proof to an \mdot proof, we need to adjust the definitions from above.

\textit{Precise typing} needs to be defined for location values.
\begin{definition}[Precise Value Typing]
  $\typPrecDftN v T$ if $\typDftN v T$ and the typing derivation of $t$ ends in 
  \rn{($\{\}$-I)}, \rn{(All-E)}, \tnew{or \rn{(LOC)}}.
\end{definition}

Since the typing relation depends on a store typing, the \textit{stack correspondence} 
relation needs to include~$\S$.
\begin{definition}[Stack Correspondence]
  A stack $\sta=\overline{x_i\mapsto v_i}$ \empty{corresponds} to 
  a type environment $\G=\overline{x_i\colon T_i}$ \tnew{and store typing $\S$},
  written $\wf\G{\new\S}\sta$, if for each $i$, $\typPrec \G {\new\S} {v_i} T$.
\end{definition}

The set of \textit{possible types} needs to include a store typing and two additional cases for references. 
First, if a value is a reference to variables of type $T$, then the reference type $\Ref T$ should be in 
the set of possible types:
if $\S(l) = T$, then $T\in\pt{\G\co\S} x l$.
Second, we need to account for reference subtyping. 
If the set of possible types includes a reference type $\Ref T$, and $U$ is both a sub- and supertype of $T$,
then $\Ref U$ is also in the set of possible types.

The updated definition of possible types is as follows.
\begin{definition}[Possible Types]
  The possible types $\pt {\G\co\new\S} x v$ of a variable $x$ bound in an environment $\G$ and
  corresponding to a value $v$ is the smallest set $\pts$ such that
  \begin{enumerate}
    \item If $v=\tNew x T d$ then $T\in\pts$.
    \item If $v=\tNew x T d$ and $\set{a=t}\in d$ and $\typDftN t {T'}$ then $\tFldDec a {T'}\in \pts$.
    \item If $v=\tNew x T d$ and $\set{A=T'}\in d$ and $\subDftN S {T'}$, $\subDftN {T'} U$ 
          then $\tTypeDec A S U\in \pts$.
    \item If $v=\tLambda x S t$ and $\typN{\extendG x S}\S t T$ and $\subDftN {S'} S$ 
          and $\subN{\extendG x {S'}}\S T {T'}$ then $\tForall x {S'}{T'}\in \pts$.
    \item \tnew{If $v=l$ and $\S(l) = T$ then $\Ref T\in\pts$.}
    \item \tnew{If $\Ref T\in\pts$, $\subDft T U$, and $\subDft U T$, then $\Ref U\in\pts$.}
    \item If $S_1\in \pts$ and $S_2\in \pts$ then $\tAnd {S_1}{S_2}\in \pts$.
    \item If $S\in\pts$ and $\typPrecDftN y {\tTypeDec A S S}$ then $y.A\in\pts$.
    \item If $T\in\pts$ then $\tRec x T\in\pts$.
  \end{enumerate}
\end{definition}

\subsection{Stores and well-typedness}
It is standard in proofs of progress and preservation to require that
an environment be well-formed with respect to a typing:
$\forall x. \G\vdash \sta(x) \colon\G(x)$. For stacks and stack typings,
this condition follows from the definition of $\G\wfsym\sta$.
We need to also define well-formedness for stores and store typings:

\noindent\tnew{\parbox{\textwidth}{
\begin{definition}[Well-Typed Store]
  A store $\sto=\set{l_i\mapsto x_i}$ is \emph{well-typed} with respect to an environment
  $\G$ 
  and store typing $\S=\overline{l_i\mapsto T_i}$, written $\wtDft$, if for each $i$,
  $\typDft{x_i}{T_i}$.
\end{definition}}}
\\[2mm]

The stronger corresponding stacks condition is not required for stores.
For stacks, it is needed to ensure absence of bad bounds, because
a type can depend on a stack variable (e.g.\ $x.A$ depends on $x$).
No similar strengthening of well-typed stores is needed because types
cannot depend on store locations.

\subsection{Proof}

In this section, we present the central lemmas required to prove the \mdot soundness theorems.

The Canonical Forms Lemma requires a well-typed store, and a statement that
values corresponding to reference types must be locations.
\begin{lemma}[Canonical Forms]
  If $\wf\G{\new{\S}}\sta$ and $\new\wtDft$, then
  \begin{enumerate}
    \item If $\typDftN x {\tForall x T U}$ then $\sta(x)=\tLambda x {T'} t$ for some $T'$ 
          and $t$ such that 
          $\subDftN T {T'}$ and $\typN{\extendG x T} \S t U$.
    \item If $\typDftN x {\tFldDec a T}$ then $\sta(x)=\tNew x S d$ for some 
          $S\co d\co t$ such that 
          $\typDftN d S$, $\set{a=t}\in d$, $\typDftN t T$.
    \item \tnew{\parbox{.94\textwidth}{
            If $\typDft x {\Ref T}$ then $\sta(x)=l$ and $\sto(l)=y$ for some $l\co y$ 
            such that $\typDft l {\Ref T}$ and $\typDft y T$.
          }}
  \end{enumerate}
\end{lemma}

The Substitution Lemma requires substitution inside of the store typing,
since the types in $\S$ can refer to 
the substituted variable.
\begin{lemma}[Substitution]
  If $\typN{\extendG x S} \S t T$ and $\typN \G {\tSubst x y \S} y {\tSubst x y S}$ then 
  $\typN \G {\tSubst x y \S} {\tSubst x y t} {\tSubst x y T}$.
\end{lemma}

The following proposition is the main soundness result of the \mdot proof. 
It is also an extension of the original proposition of the WadlerFest DOT soundness proof.
\begin{proposition}\label{eq:prop}
  Let
  \begin{itemize}
    \item $\typDftN t T$, 
    \item $\wf\G{\new\S}\sta$, and 
    \item$\new{\wtDft}$.
  \end{itemize} Then either
  \begin{itemize}
    \item $t$ is an answer, or
    \item there exist a stack $\sta'$, \tnew{store $\sto'$} and a term $t'$ such that
          $\reductionFull{\sst\sta {\new{\sto}} t}{\sst {\sta'} {\new{\sto'}} {t'}}$ and 
          for any 
          such $\sta'\co\new{\sto'}\co t'$
          there exist environments $\G'$ and $\new{\S'}$ such that
          \begin{itemize}
            \item $\typN{(\G\co\G')}{(\S\co\S')}{t'}T$, 
            \item $\wf{(\G\co\G')}{\new{(\S\co\S')}}\sta$, and
            \item $\new{\wt{(\G\co\G')}{(\S\co\S')}\sto}$.
          \end{itemize}
  \end{itemize}
\end{proposition}
Progress and preservation (Theorems~\ref{thm:progress} and~\ref{thm:preservation}) follow
directly from Proposition~\ref{eq:prop}, if we assume $\G$ to be empty.

\section{Discussion}\label{sec:discussion}

In this section, we explain the design choices of \mdot in more detail and discuss
possible alternative implementations.

\subsection{Motivation for a store of variables}\label{sec:heapmotivation}
One unusual aspect of the design of \mdot is that the store contains
variables rather than values. We experimented with alternative designs
that contained values, and observed the following undesirable
interactions with the existing design of WadlerFest DOT.

A key desirable property is that the store should be well-typed
with respect to a store typing:
$\forall l.\ \G,\S\vdash \sto(l) : \S(l)$.

\newcommand{\fatop}{\tFldDec a \top}
\newcommand{\newvalue}{\tNew y \fatop {\{a = t\}}}
Many of the WadlerFest DOT type assignment rules apply only to variables, and not
to values. For example, the type $\fatop$ is not inhabited by any
value, but a variable can have this type.
This is because an object
value has a recursive type, and the (\rn{Rec-E}) rule that opens a
recursive type $\tRec x \fatop$ into $\fatop$ applies only to variables, not to values.
In particular, in the term
$$\tLet x \newvalue \tRef x \fatop$$
$x$ has type $\fatop$ but $\newvalue$ does not,
even though the let binding suggests that the variable
and the value should be equal. If memory cells were to contain
values, a cell of type $\tFldDec a \top$ would not make
sense, because no values have that type. However, since WadlerFest DOT
prohibits subtyping between recursive types, this would
severely restrict the polymorphism of memory cells.
In particular, it would be impossible to define a memory cell
containing objects with a field $a$ of type $\top$
and possibly additional fields. Extending the subtyping
rules to apply to values as well as variables would disrupt the delicate WadlerFest DOT soundness proofs.

The above example $\sf{let}$ term demonstrates another problem: type preservation.
The type system should admit the term 
$\tRef x \fatop$ because $x$ has type $\fatop$.
This term should reduce to a fresh location $l$ of type $\Ref \fatop$.
But a store that maps $l$ to $\newvalue$ would not be well typed,
because the value does not have type $\fatop$.

\subsection{Correctness of a store of variables}\label{sec:heapcorrectness}
Putting variables instead of values in the store
raises a concern: when we write a variable into the
store, we expect that when we read it back, it will
still be in scope, and it will still be bound to the
same value.
For example, in the following program fragment, the variable $x$ gets saved in the store
inside the function $f$.
\[
  \begin{matrix*}[l]
    &\mlet f&=&\tLambda x \top {\tRef x T}&\tin\\
    &\mlet y&=&v&\tin\\
    &\mlet r&=&f\,y&\tin\\
    &!r
  \end{matrix*}
\]
Will $x$ go out of scope by the time we read it from the store?

\begin{figure*}
\[
  \begin{matrix*}[l]
      \emptyset & \mid & f \mapsto {\tLambda x \top {\tRef x T}},\ y \mapsto v & \mid & \tLet r {f\,y} {!r}                           & \red\\
      \emptyset & \mid & f \mapsto {\tLambda x \top {\tRef x T}},\ y \mapsto v & \mid & \new{\tLet r {\tSubst x y {\tRef x T}} {!r}}  & \red\\
      \emptyset & \mid & f \mapsto {\tLambda x \top {\tRef x T}},\ y \mapsto v & \mid & \new{\tLet r {\tRef y T} {!r}}                & \red\\
      \new{l \mapsto y} & \mid & f \mapsto {\tLambda x \top {\tRef x T}},\ y \mapsto v & \mid & \new{\tLet r l {!r}}  & \red\\
      l \mapsto y & \mid & f \mapsto {\tLambda x \top {\tRef x T}},\ y \mapsto v,\ \new{r \mapsto l} & \mid & \new{!r}& \red\\
      l \mapsto y & \mid & f \mapsto {\tLambda x \top {\tRef x T}},\ y \mapsto v,\ r \mapsto l & \mid & \new y\\
  \end{matrix*}
\]
\caption{Reduction sequence for example program}
\label{fig:redeg}
\end{figure*}

The reduction sequence for this program is shown in Figure~\ref{fig:redeg}.
Notice that before the body $\tRef x T$ of the function is reduced,
the parameter $x$ is first substituted with the argument $y$, which does not
go out of scope.

More generally, from the stack-based reduction semantics in Figure~\ref{fig:red2},
it is immediately obvious that when a variable $x$ is saved in the store
using $\tRef x T$ or $\tAsgn y x$, the only variables that are in scope
are those on the stack. There are no function parameters in scope
that could go out of scope when the function finishes.

Moreover, once a variable is on the stack, it never goes out of scope,
and the value that it is bound to never changes. This is because the only
reduction rule that modifies the stack is (\rn{Let-Value}), and it only
adds a new variable binding, but does not affect any existing bindings.

\subsection{Creating references}\label{sec:refs}

The \mdot reference creation term $\tRef x T$ requires both a type $T$
and an initial variable $x$. The variable is needed so that a reference
cell is always initialized, to avoid the need to add a \textsf{null} value to DOT.
If desired, it is possible to model uninitialized memory cells in \mdot by explicitly
creating a sentinel null value.

Some other calculi with mutable references (e.g. Types and Programming Languages~\citep{Pierce:2002:TPL:509043})
do not require the type $T$ to be given explicitly, but just adopt the precise
type of $x$ as the type for the new cell. Such a design does not fit well
with subtyping in DOT. In particular, it would prevent the creation of a
cell with some general type $T$ initialized with a variable $x$ of a more
specific subtype of $T$.

More seriously, such a design (together with subtyping) would break type preservation.
Suppose that $\typDft y S$ and $\subDft S T$.
Then we could arrive at the following reduction sequence:
{\setlength{\mathindent}{0cm}\begin{align*}
  \emptyset &&& \mid & f\mapsto \tLambda x T {\code{ref } x}\co y\mapsto v &&& \mid & f\,y                       && \red\\
  \emptyset &&& \mid & f\mapsto \tLambda x T {\code{ref } x}\co y\mapsto v &&& \mid & \new{\tSubst x y {\code{ref }x}} && \red\\
  \emptyset &&& \mid & f\mapsto \tLambda x T {\code{ref } x}\co y\mapsto v &&& \mid & \new{\code{ref }y}
\end{align*}}
The term at the beginning of the reduction sequence has type $\Ref T$, while the term at the end,
$\code{ref }y$, has type $\Ref S$.
Preservation would require $\Ref S$ to be a subtype of $\Ref T$,
but this is not the case in general since the only condition that this
example imposes on $S$ and $T$ is that $\subDft S T$.

\section{Example}\label{sec:example}
Recall the aquarium example from Section~\ref{sec:dot}. 
Suppose we wanted to make the aquarium mutable:
instead of returning a new \textsf{Aquarium},
the \textsf{addFish} method
should update the aquarium's list of fish by appending the new fish
object to it. 
A possible implementation in \mdot looks as follows:
  \listingNoCapt{ML}{mutable-aquarium.dot}
The \textsf{fish} member of the new \textsf{Aquarium} version is now a \textit{reference}
to a list of fish, and the \textsf{addFish} changes the list to include the new fish and returns it.

\section{Related Work}\label{sec:related}

The semantics of mutable references presented in this paper is similar to Pierce's 
extension of the simply-typed lambda calculus with typed mutable
references~\citep[Chapter~13]{Pierce:2002:TPL:509043}.
However, the resemblance is mostly syntactic: the language presented in the book
does not include subtyping or other object-oriented features.

An extensive study of object-oriented calculi, including ones that support mutation, can be found in
``A Theory of Objects''~\citep{DBLP:books/daglib/0084624}.
The book surveys imperative calculi with a range of advanced object-oriented features,
including subtyping and inheritance, self types, and typed mutable objects (using
protected storage cells).
\citet{DBLP:conf/ecoop/MackayMPGC12} developed a version of Featherweight
Java~\citep{fj} with mutable and immutable objects and formalized it in Coq.
However, neither of the analyzed type systems involved path-dependent
types.

The $\nu$Obj calculus~\citep{nuobj} introduced types as members of objects,
and thus path-dependent types. However, type members had only upper bounds,
but not lower bounds, as they do in Scala. On the other hand, the $\nu$Obj
calculus was richer than DOT, including features such as first-class
classes, which are not present even in the full Scala language.
Featherweight Scala~\citep{fs} was a simpler calculus intended to
correspond more closely to Scala, and with decidable type-checking.
However, its type system has not been proven sound.
A related calculus, Scalina~\citep{scalina}, intended to explore
the design of higher-kinded types in Scala, was also not proven sound.

\citet{fool12} first used the name DOT for
a calculus intended to be simple, and to capture only essential features,
namely path-dependent types, type refinement, intersection, and union.
This paper discussed the difficulties with proving such a calculus
sound. The most notable challenge were counterexamples to type preservation
in a small-step semantics. In general, a term can reduce to another term
with a narrower type. In this DOT calculus, this narrowing could disrupt
existing subtyping relationships between type members in that type.

\citet{oopsla14} examined simpler calculi with subsets of the features
of DOT to determine which features cause type preservation to fail.
They identified the problem of bad bounds, noted that they cannot occur
in runtime objects that are actually instantiated, and conjectured
that distinguishing types realizable at runtime could lead to a
successful soundness proof for a DOT calculus with all of its features.
\citet{arxiv1} confirmed this conjecture by providing the first
soundness proof of a big-step semantics for a DOT calculus with type
refinement and a type lattice with union and intersection. The use of
a big-step semantics makes it possible to get around the problem of
small steps temporarily violating type preservation, at the cost of
a more complex soundness proof.
An update to the technical report~\citet{arxiv2} reports that the authors were also able to add
mutable references to this big-step version of the calculus.

WadlerFest DOT~\citep{Amin2016} defines a very specific evaluation order for the
subexpressions of a DOT calculus that satisfies type preservation at
each reduction step, and expressed it in a small-step semantics. The
semantics uses administrative normal form (ANF) to make the necessary
evaluation order explicit and clear, and to distinguish realizable types of
objects instantiated at run time from arbitrary types. In particular,
in the context in which a term is reduced, every ANF variable maps to a
value, an actual run-time object, rather than an arbitrary term; thus,
the ANF variables play the role of labels of run-time values in the
semantics and its proof. The paper is accompanied by a Coq formalization
of the full type soundness proof in the familiar style of progress and
preservation~\cite{felleisen}, and is thus well suited as a basis for extensions to the
calculus. It is this WadlerFest DOT calculus that we
have extended with mutable references, to serve as a basis for further
extensions that involve mutation.

One limitation of the WadlerFest DOT calculus is the lack of subtyping
between recursive types. The calculus of \citet{oopsla16}, which is to
appear, will remove this limitation, while maintaining a small step
semantics. We hope that the experience that we have reported here
on the WadlerFest DOT calculus will also be helpful for adding mutation
to this new DOT calculus.

\section{Conclusion}

WadlerFest DOT formalizes the essence of Scala, but it lacks mutation, which is an important feature of object-oriented languages.
In this paper, we show how WadlerFest DOT can be extended to handle mutation in a type-safe way.

As shown in the paper, adding a mutable store to the semantics of WadlerFest DOT is not
straightforward.
The lack of subtyping between recursive
types leads to situations where variables and values, even though they are
bound together, have incompatible types.
As a result, if WadlerFest DOT were extended with a conventional store containing
values, it would be impossible for a cell of a given type $T$ to store values of
different subtypes of $T$,
thus significantly restricting the kinds of mutable code that could be expressed.

The key idea of this paper is to enable support for mutation in WadlerFest DOT by using
a store that contains variables instead of values.
We have shown that by using a store of variables, it is possible to extend WadlerFest DOT with
mutable references in a type-safe way.
This leads to a formalization of a language with path-dependent types
and mutation, and also brings WadlerFest DOT one step closer to encoding the full Scala
language.

\paragraph{Acknowledgement:} This research was supported by the Natural Sciences and Engineering Research Council of Canada.

\bibliography{bibliography}

\begin{thebibliography}{19}
\providecommand{\natexlab}[1]{#1}
\providecommand{\url}[1]{\texttt{#1}}
\providecommand{\urlprefix}{}

\bibitem[{Abadi and Cardelli(1996)}]{DBLP:books/daglib/0084624}
Abadi, M., Cardelli, L.: A Theory of Objects.
\newblock Monographs in Computer Science, Springer (1996)

\bibitem[{Amin(2016{\natexlab{a}})}]{aminphd}
Amin, N.: Dependent Object Types.
\newblock Ph.D. thesis (2016{\natexlab{a}})

\bibitem[{Amin(2016{\natexlab{b}})}]{bugrep}
Amin, N.: Soundness issue with path-dependent type on \textsf{null} path.
\newblock \url{https://issues.scala-lang.org/browse/SI-9633}
  (2016{\natexlab{b}})

\bibitem[{Amin et~al.(2016)Amin, Gr{\"u}tter, Odersky, Rompf, and
  Stucki}]{Amin2016}
Amin, N., Gr{\"u}tter, S., Odersky, M., Rompf, T., Stucki, S.: The Essence of
  Dependent Object Types.
\newblock Springer International Publishing, Cham (2016)

\bibitem[{Amin et~al.(2012)Amin, Moors, and Odersky}]{fool12}
Amin, N., Moors, A., Odersky, M.: Dependent {O}bject {T}ypes.
\newblock In: International Workshop on Foundations of Object-Oriented
  Languages (FOOL 2012) (2012)

\bibitem[{Amin et~al.(2014)Amin, Rompf, and Odersky}]{oopsla14}
Amin, N., Rompf, T., Odersky, M.: Foundations of path-dependent types.
\newblock In: Proceedings of the 2014 {ACM} International Conference on Object
  Oriented Programming Systems Languages {\&} Applications, {OOPSLA} 2014, part
  of {SPLASH} 2014, Portland, OR, USA, October 20-24, 2014. pp. 233--249 (2014)

\bibitem[{Amin and Tate(2016)}]{null}
Amin, N., Tate, R.: Java and scala's type systems are unsound: The existential
  crisis of null pointers.
\newblock In: to appear in OOPSLA 2016 (2016)

\bibitem[{Cremet et~al.(2006)Cremet, Garillot, Lenglet, and Odersky}]{fs}
Cremet, V., Garillot, F., Lenglet, S., Odersky, M.: A core calculus for {Scala}
  type checking.
\newblock In: Mathematical Foundations of Computer Science, 31st International
  Symposium, Slovakia (2006)

\bibitem[{Igarashi et~al.(2001)Igarashi, Pierce, and Wadler}]{fj}
Igarashi, A., Pierce, B.C., Wadler, P.: Featherweight {Java}: a minimal core
  calculus for {Java} and {GJ}.
\newblock {ACM} Trans. Program. Lang. Syst. 23(3), 396--450 (2001)

\bibitem[{Mackay et~al.(2012)Mackay, Mehnert, Potanin, Groves, and
  Cameron}]{DBLP:conf/ecoop/MackayMPGC12}
Mackay, J., Mehnert, H., Potanin, A., Groves, L., Cameron, N.R.: Encoding
  {F}eatherweight {J}ava with assignment and immutability using the {C}oq proof
  assistant.
\newblock In: Proceedings of the 14th Workshop on Formal Techniques for
  Java-like Programs (2012)

\bibitem[{Moors et~al.(2008)Moors, Piessens, and Odersky}]{scalina}
Moors, A., Piessens, F., Odersky, M.: Safe type-level abstraction in scala.
\newblock In: International Workshop on Foundations of Object-Oriented
  Languages (FOOL 2008) (2008)

\bibitem[{Odersky(2016)}]{sdts}
Odersky, M.: Scaling {DOT} to {S}cala --- {S}oundness.
\newblock
  \url{http://www.scala-lang.org/blog/2016/02/17/scaling-dot-soundness.html}
  (2016)

\bibitem[{Odersky et~al.(2003)Odersky, Cremet, R\"ockl, and Zenger}]{nuobj}
Odersky, M., Cremet, V., R\"ockl, C., Zenger, M.: A nominal theory of objects
  with dependent types.
\newblock In: Proc. ECOOP'03. Springer LNCS (2003)

\bibitem[{Petrashko(2016)}]{init}
Petrashko, D.: Making sense of initialization order in scala.
\newblock \url{https://d-d.me/talks/scalar2016/#/} (2016)

\bibitem[{Pierce(2002)}]{Pierce:2002:TPL:509043}
Pierce, B.C.: Types and Programming Languages.
\newblock The MIT Press, 1st edn. (2002)

\bibitem[{Rompf and Amin(2015)}]{arxiv1}
Rompf, T., Amin, N.: From {F} to {DOT:} type soundness proofs with definitional
  interpreters.
\newblock CoRR abs/1510.05216v1 (2015),
  \urlprefix\url{http://arxiv.org/abs/1510.05216v1}

\bibitem[{Rompf and Amin(2016{\natexlab{a}})}]{arxiv2}
Rompf, T., Amin, N.: From {F} to {DOT:} type soundness proofs with definitional
  interpreters.
\newblock CoRR abs/1510.05216v2 (2016{\natexlab{a}}),
  \urlprefix\url{http://arxiv.org/abs/1510.05216v2}

\bibitem[{Rompf and Amin(2016{\natexlab{b}})}]{oopsla16}
Rompf, T., Amin, N.: Type soundness for dependent object types ({DOT}).
\newblock In: to appear in OOPSLA 2016 (2016{\natexlab{b}})

\bibitem[{Wright and Felleisen(1994)}]{felleisen}
Wright, A.K., Felleisen, M.: A syntactic approach to type soundness.
\newblock Inf. Comput. 115(1), 38--94 (1994)

\end{thebibliography}


\begin{thebibliography}{0}
\providecommand{\natexlab}[1]{#1}
\providecommand{\url}[1]{\texttt{#1}}
\expandafter\ifx\csname urlstyle\endcsname\relax
  \providecommand{\doi}[1]{doi: #1}\else
  \providecommand{\doi}{doi: \begingroup \urlstyle{rm}\Url}\fi

\end{thebibliography}

\end{document}